\documentclass[journal abbreviation, manuscript]{copernicus}

\begin{document}

\title{Constructing Extreme Heatwave Storylines with Differentiable Climate Models}


\Author[1][whittaker.tim@courrier.uqam.ca]{Tim}{Whittaker} 
\Author[1]{Alejandro}{Di Luca}

\affil[1]{Centre Étude et simulation du climat à l'échelle régionale (ESCER), D\'epartement des Sciences de la Terre et de l’Atmosph\`ere, Universit\'e du Qu\'ebec \`a Montr\'eal, Montr\'eal, Qu\'ebec, Canada}

\runningtitle{ }

\runningauthor{ }

\received{}
\pubdiscuss{} 
\revised{}
\accepted{}
\published{}


\firstpage{1}

\maketitle

\nolinenumbers

\begin{abstract}
Understanding the plausible upper bounds of extreme weather events is essential for risk assessment in a warming climate. Existing methods, based on large ensembles of physics-based models, are often computationally expensive or lack the fidelity needed to simulate rare, high-impact extremes. Here, we present a novel framework that leverages a differentiable hybrid climate model, NeuralGCM, to optimize initial conditions and generate physically consistent worst-case heatwave trajectories. Applied to the 2021 Pacific Northwest heatwave, our method produces heatwave intensity up to 3.7 $^\circ$C above the most extreme member of a 75-member ensemble. These trajectories feature intensified atmospheric blocking and amplified Rossby wave patterns—hallmarks of severe heat events. Our results demonstrate that differentiable climate models can efficiently explore the upper tails of event likelihoods, providing a powerful new approach for constructing targeted storylines of extreme weather under climate change.
\end{abstract}

\introduction  
\label{intro}

The 2021 Pacific Northwest (PN2021) heatwave shattered historical temperature records, culminating in Lytton, Canada’s unprecedented 49.6$^\circ$C observation—a 4.6$^\circ$C increase over the country’s previous record measurement \citep{White2023, ThePacificNorthwestHeatWave}. This event, virtually implausible under preindustrial conditions \citep{esd-13-1689-2022}, exemplifies a critical challenge in climate science: determining the upper bounds of what is physically possible for different weather extremes under current or future climatic conditions. 

The PN2021 heatwave emerged from persistent atmospheric blocking sustained by large-scale Rossby waves that disrupted zonal flow and stalled a high-pressure system over the region \citep{ThePacificNorthwestHeatWave, White2023}. This large-scale setup was fueled by upstream dynamics. \cite{Mo2022} linked it to anomalous atmospheric river activity, while \citet{https://doi.org/10.1029/2021GL097699} identified that diabatic heating within the warm conveyor belt of an upstream cyclone provided the necessary Rossby wave activity to establish the block. Once established, the block suppressed cloud formation and drove prolonged subsidence, adiabatically warming near-surface air masses \citep{MeteorologicalAnalysisofthePacificNorthwestJune2021Heatwave}. \cite{White2023} corroborated the importance of these mechanisms and estimated via four-day backward trajectory analysis that diabatic processes accounted for approximately 78\% of the net temperature change of air parcels entering the region, with the remaining $\sim 22\%$ attributed to adiabatic warming from subsidence. Locally, dry soil conditions further intensified these temperatures through non-linear land-atmosphere interactions \citep{Bartusek2022, TheInfluenceofSoilMoistureontheHistoric2021PacificNorthwestHeatwave, https://doi.org/10.1029/2022EF002967}. By studying a 100-member ensemble of PN2021 with varying initial land surface conditions, \cite{https://doi.org/10.1029/2025EF006216} found that variations in antecedent soil moisture led to a spread of approximately $3^\circ$C in peak temperatures, largely driven by regions shifting into a transitional evaporation regime where latent heat flux becomes highly sensitive to soil moisture.

To systematically explore such extremes, storylines are increasingly used, representing physically consistent sequences of weather events that depict how a counterfactual extreme event might occur \citep{Hazeleger2015, Shepherd2019, https://doi.org/10.1029/2020EF001783}. This approach enables a mechanistic exploration of how minor perturbations can lead to the amplification of extreme events. Here, we use a novel differentiable modeling framework to demonstrate that targeted initial-condition perturbations can further amplify these typical extreme trajectories, giving extreme heatwave storylines.

Identifying storylines for the most extreme weather events is a needle-in-a-haystack problem due to their inherent rarity. The traditional approach is the use of single-model initial-condition large ensembles \citep{Deser2020, Suarez-Gutierrez2020, esd-12-401-2021, Diffenbaugh_2021}, and more recently, so-called huge ensembles \citep{mahesh2024hugeensemblesidesign, mahesh2024hugeensemblesiiproperties}, in which vast numbers of model runs allow the exploration of a wide range of potential outcomes. By systematically increasing ensemble size, the chances of capturing low-probability extremes increase. However, these ensembles are computationally demanding and not very effective at sampling the full range of outcomes. In addition, due to their high computational cost, it is virtually impossible to perform such ensembles using kilometer-scale simulations, which are required to well simulate some types of weather extreme events (e.g., extreme convective precipitation). 

In recent years, a number of approaches have been proposed to generate extreme event storylines \citep{doi:10.1073/pnas.1712645115, https://doi.org/10.1029/2018MS001419, 10.1063/1.5081461, gmd-13-763-2020, VeryRareHeatExtremesQuantifying, Fischer2023}. These methods focus computing resources on specific extreme events, instead of continuous long simulations. Some approaches enhance the likelihood of simulating extreme events by constructing targeted ensembles \citep{doi:10.1073/pnas.1712645115,10.1063/1.5081461,Fischer2023}. \cite{Fischer2023} focus on generating an initial condition ensemble of climate model simulations of known extreme events using a method named ensemble boosting. They applied this approach to the PN2021 heatwave and by perturbing the initial conditions using numerical noise for 500 members, they found a 5-day running average of daily maximum temperature anomalies up to 2.9$^\circ$C larger than the unperturbed event. Other approaches to construct storylines of extreme events use the large deviation algorithm \citep{doi:10.1073/pnas.1712645115,https://doi.org/10.1029/2020GL091197, Statisticalanddynamicalaspectsofextremelyhotsummersin} where an ensemble of simulations is ran and members are periodically pruned or cloned such that an ensemble most likely to lead to an extreme event is generated. Focusing on western European heatwaves and using the large deviation algorithm, \cite{doi:10.1073/pnas.1712645115} generated an ensemble which has a mean 2$^\circ$C anomaly compared to a control ensemble of 128 members. Other applications of the algorithm showed its ability to identify even more extreme events, with ensembles with mean anomalies of 4$^\circ$C \citep{https://doi.org/10.1029/2020GL091197}. Meanwhile, \cite{https://doi.org/10.1029/2018MS001419} introduced a variational data assimilation technique, optimized with a 4D-Var inspired method to intensify past extreme tropical cyclones with minimal perturbations. This approach is closely related to the method presented here; however, we leverage automatic differentiation and computationally efficient ML-based models. 

Recent advances in machine learning (ML) have led to the development of transformative tools for weather and climate modeling. Neural network architectures like GraphCast \citep{lam2023graphcastlearningskillfulmediumrange}, Pangu-Weather \citep{Bi2023}, FourCastNet \citep{pathak2022fourcastnetglobaldatadrivenhighresolution, 10.1145/3592979.3593412}, and FuXi \citep{Chen2023} have demonstrated forecasting skill comparable to that of traditional numerical weather prediction systems, but at significantly reduced computational costs \citep{https://doi.org/10.1029/2023MS004019, ValidatingDeepLearningWeatherForecasts, ennis2025turningheatassessing2m, zhang2025numericalmodelsoutperformai}. In addition to their reduced computational costs, these models by construction allow us to define optimization problems on them that can be solved through gradient-based optimizers. \cite{https://doi.org/10.1029/2023MS004019} introduced a standardized benchmark to compare the various ML models against ERA5 and the European Centre for Medium-Range Weather Forecasts's (ECMWF) integrated forecast system (IFS). Using this benchmark, it is shown that deterministic, data-driven methods such as Pangu-Weather, GraphCast, and FuXi result in similar root-mean-square error (RMSE) in forecasting near-surface temperature, wind, and pressure up to ten days ahead. However, their forecast skill deteriorates rapidly for longer lead times, resulting in overly smoothed predictions. 
Using three case studies, \cite{ValidatingDeepLearningWeatherForecasts} evaluated GraphCast, Pangu-Weather and FourCastNet against ERA5 reanalysis and ECMWF’s IFS for the PN2021 heatwave, the 2023 South Asian humid heatwave and a 2021 North American winter storm. They find that all data-driven models systematically underestimate the peak 2m temperature during the PN2021 heatwave, with root mean square error (RMSE) values at grid points near Vancouver, Seattle, and Portland exceeding twice the 10-day IFS error and reaching up to four times that value in Portland. During the South Asian humid heatwave, data-driven forecasts of heat index computed from 2-m air temperature and 1000-hPa relative humidity underpredicted observed peaks more strongly than IFS, particularly over Bangladesh. For the North American winter storm, data-driven forecasts of wind chill at College Station, Texas, achieved lower peak errors than IFS, with Pangu-Weather and GraphCast outperforming the operational model.

An alternative to purely data-driven approaches is the use of hybrid models, such as NeuralGCM \citep{Kochkov2024}, which combines a traditional dynamical core with ML components. \citep{duan2024aibasedclimatemodelevaluation} have shown NeuralGCM’s ability to hindcast the PN2021 heatwave, though due to the lack of processes (such as land-atmosphere feedbacks), the intensity of the heatwaves tends to be underestimated. Similarly to the purely data-driven models, NeuralGCM produces surface variables forecasts with skill comparable to that of the ECMWF IFS system \cite{https://doi.org/10.1029/2023MS004019}. Moreover, the use of the dynamical core both prevents the evolved fields from being overly smoothed and enhances numerical stability. These benefits allow for longer time integrations and make the model suitable for climate studies \citep{Kochkov2024}. 

These new types of models are by construction differentiable through automatic differentiation \citep{gmd-16-3123-2023}. The automatic differentiation property enables efficient optimization, allowing gradient-based exploration storylines in high-dimensional climate models. This is in line with many new extreme event opportunities enabled by ML models \citep{https://doi.org/10.1002/wcc.914,Camps-Valls2025}. Leveraging automatic differentiation, recent studies have implemented variational data assimilation techniques using neural networks, with applications ranging from toy models, such as the Lorenz 96 system \citep{75462}, to reduced-order physical representations of the atmosphere \citep{solvik20244dvarusinghessianapproximation,manshausen2024generativedataassimilationsparse}. Additionally, \cite{vonich2024predictabilitylimit2021pacific} demonstrated that the differentiability of GraphCast allows for a more accurate reconstruction of the initial conditions that led to the PN2021 heatwave compared to using ERA5 reanalysis data. \cite{Baño-Medina2025} explores the use of ML models and automatic differentiation to perform sensitivity analysis of the initial conditions leading to the development of cyclone Xynthia. Their findings suggest that gradients computed from the data-driven weather model at a 36-hour lead time exhibit sensitivity structures that closely resemble those generated by the adjoint of a dynamical model. In other words, the evolved perturbations from both approaches lead to similar impacts on the cyclone’s evolution. 

In this study, we focus on the PN2021 heatwave event due to its well-documented synoptic drivers, and its prevalence in extreme event studies \citep{Lucarini_2023, Fischer2023,White2023,esd-13-1689-2022}. We use the automatic differentiation feature of the NeuralGCM model to optimize perturbed initial conditions, and we identify trajectories where enhanced geopotential height anomalies intensify downstream near-surface temperature extremes. These storylines reveal heatwave intensity increases of $3.7^\circ$C beyond the extreme temperatures obtained from a 75-member ensemble run using NeuralGCM for the event, analogous to the ensemble boosting approach \citep{Fischer2023}. Our results demonstrate the potential of differentiable hybrid models for investigating worst-case scenarios, offering a computationally efficient alternative to traditional, computationally expensive, large ensembles. 

This paper is organized as follows: Section \ref{Sec:Methods} introduces an optimization problem whose minimization yields extreme heatwaves and describes how NeuralGCM is used to solve it. Section \ref{Sec:Results} presents the optimized heatwave storylines in comparison with an ensemble run. Section \ref{Sec:Discussion} discusses the implication of the method and future directions. Finally, Section \ref{Sec:Conclusions} concludes the paper.

\section{Methods}
\label{Sec:Methods}

\subsection{Initial Conditions Optimization Problem}
\label{subsec:optimization}
Our goal is to find the worst-case physically plausible heatwave trajectory our model can produce. To achieve this, we must find the specific, small perturbations to a known initial state that will evolve into the most extreme event. This search is formulated as an optimization problem, where we define a loss function that the model will automatically minimize in an iterative way to find these optimal initial-state perturbations. Formally, a suitable loss function for our problem is one that
\begin{enumerate}
    \item maximizes a target extreme event, and
    \item minimizes the introduced perturbation.
\end{enumerate}

The optimization process is framed through a continuous-time dynamical system. For conceptual clarity, we describe the problem using an ordinary differential equation system:
\begin{equation}
    \dot{\vec{x}} = b(\vec{x}(t,\vec{x}_0)),
\end{equation}
with $b$ representing some nonlinear operator, $t$ the time and $\vec{x}_0$ some initial state $t_0$. The solution is given by $\vec{X}(t) = S_t \vec{x}_0$ where $S_t$ is the evolution operator up to time $t$. The core aim of the optimization problem is to identify the initial conditions $\vec{x}_0$ that drive the dynamical system toward an extreme desired state, represented by a target observable $\mathcal{O}(\vec{X}(t))$. Given a baseline initial state $\vec{x}_0^b$, we define a perturbation $\Delta \vec{x}_0 = \vec{x}_0^b - \vec{x}_0$. The optimization problem is formulated in terms of minimizing the following loss function:
\begin{equation}
    \mathcal{L}(\vec{X}(t), \Delta \vec{x}_0) = F(\mathcal{O}(\vec{X}(t))) + \vec{\lambda} \cdot \Delta \vec{x}_0^2,
\end{equation}
where the first term is designed to favor more extreme values of the observable by applying a cost function $F$ to the outcome $\mathcal{O}(\vec{X}(t))$, and the second term, scaled by the regularization parameter $\vec{\lambda}$, penalizes the magnitude of the initial perturbation $\Delta \vec{x}_0^i$. This formulation balances the competing objectives of inducing a rare event and keeping the initial perturbation sufficiently small.

In particular, we pick our observable $(\mathcal{O}(\vec{X}(t)))$ to be the temperature over a domain $\mathcal{D}$ and over a period of time $\tau$ at the 1000-hPa pressure level of the model ($\frac{1}{\tau |D|}\int_{0}^{\tau}\int_{D} T_{1000}(\phi, \theta,t)\,d\phi \,d\theta \,dt$), with $\phi, \theta$ being the longitude and latitude. Multiple functions $F(\mathcal{O}(\vec{X}$) can be considered, but our main results uses $F(X) = \frac{c}{X}$ which gives us the loss :
\begin{equation}
\label{eq:opt_problem}
\mathcal{L}(T_{1000}, \Delta x_0) 
= \underbrace{\beta\frac{T_{\text{ref}}}{\frac{1}{\tau |D|}\int_{0}^{\tau}\int_{D} T_{1000}(\phi, \theta,t)\,d\phi \,d\theta \,dt}}_{\text{Temperature objective term}} 
+ \underbrace{\sum_{i}\lambda_i\frac{(\Delta x_{0,i})^2}{(\Delta x_{\text{ref},i})^2}}_{\text{Perturbation penalty term}}
\end{equation}

where $\mathcal{D}$ corresponds to the region shown in Fig.~\ref{fig:timeseries_main_var}, and $\tau$ is set to 5 days. This 5-day period was chosen to fully encompass the three peak days of the PN2021 event, with a two-day buffer at the end, which we found aided in optimization. The terms in the loss function are normalized by their initial means, with $T_{\text{ref}}$ representing a characteristic temperature scale and $\Delta x_{\text{ref},i}$ denoting a reference perturbation scale for each perturbed variable $i = \{$Temperature, Surface Pressure, Vorticity, Divergence, Specific Humidity, Specific Cloud Ice Water Content, Specific Cloud Liquid Water Content$\}$. The normalization scale for each perturbed variable, $\Delta x_{\text{ref},i}$, is defined as the absolute mean of each respective initial field. This ensures each terms is of a similar magnitudes.

Once the simulation are optimized, we  evaluate their success through an intensity metric for the heatwaves. We define a heatwave event as a period during which the daily temperature exceeds the 99th percentile threshold for consecutive days \citep{SP3/Z4Y0LK_2023}. This definition relies on the persistence of temperature extremes (see also heatwave intensity definition); if the temperature drops below the threshold for even a single day, the event is considered terminated, and any subsequent exceedances are treated as distinct, separate events. The intensity of the heatwave is measured by the average exceedance of the temperature above the threshold over the duration of the event. Specifically, if $L$ denotes the length of the event, $T_i$ the mean temperature timeseries over a region, and $T_{\text{thresh}}$ the 99th percentile threshold, then the intensity $I$ is defined as
\begin{equation} 
\label{eq:hw_def}
I = \frac{1}{L} \sum_{i=1}^{L} (T_i - T_{\text{thresh}}). 
\end{equation}
In our analysis we compare the intensity, $I$, of the heatwaves from the optimized runs with those from the ensemble runs over the targeted five days of the optimization process.

\subsection{Numerical Implementation using NeuralGCM}

To simulate the dynamics and evaluate the loss function, we use the NeuralGCM model \citep{Kochkov2024}. Most of the experiments are performed with a horizontal grid spacing of $2.8^\circ$ (denoted as NeuralGCM2.8) because it is more computationally tractable and a climate simulation is available at this resolution. For sensitivity analysis we also consider simulations performed using a horizontal grid spacing of $1.4^\circ$. NeuralGCM employs a dynamical core to solve the primitive equations using a semi-implicit time-integration scheme and a spectral method. Physical processes on the other hand are emulated by learned physics through a neural network.

NeuralGCM has been implemented in JAX \citep{jax2018github} and supports automatic differentiation. This enables the computation of gradients with respect to both initial conditions and internal system parameters, facilitating backpropagation through the physical dynamics and neural network components. In this work, we compute gradients only with respect to the initial variables involved in the dynamical core of NeuralGCM, keeping all other parameters fixed. The loss, as defined in Eq. \ref{eq:opt_problem}, is minimized using gradient descent, specifically with the Adam optimizer from Optax \citep{deepmind2020jax}. The optimal perturbations are applied on the spherical harmonic coefficients representation of the variables. We choose NeuralGCM over other possible models because, it has demonstrated competitive forecast skill for temperatures up to 10-day lead times, contains a dynamical core and relies on a single initial condition.

Although the model runs efficiently on a single GPU with relatively low memory requirements, gradient computation demands substantial memory, scaling rapidly with the number of time steps. To address this, we employ gradient checkpointing and chunking strategies to manage memory usage. These techniques store only essential intermediate values during the forward pass, recomputing them during the backward pass to reduce memory overhead \citep{Kochkov2024}. The optimization scheme on the $2.8^\circ$ model runs on a 16GB A4000 Nvidia GPU, whereas the $1.4^\circ$ model necessitates a 40GB A100 Nvidia GPU.

We investigate extreme events by perturbing the initial conditions primarily around PN2021 using data from the ERA5 reanalysis dataset \citep{ERA5}.  We conducted two independent optimization experiments, hereafter referred to as ``EXP50'' and ``EXP75''. Their configurations—including the learning rate ($\alpha$), loss-function weights ($\beta$, $\lambda_i$), forecast lead times, initialization dates, and number of gradient descent steps (N)—are detailed in Table \ref{tab:params}. These parameters were selected via an experimental approach analogous to machine learning hyperparameter tuning, as an exhaustive automated search would be computationally prohibitive. We initially selected $N=75$ to establish a baseline comparable in computational cost to a 75-member ensemble. Subsequently, we performed the $N=50$ experiment to assess whether similar results could be achieved with fewer resources. This required retuning the $\lambda_i$ parameters; generally, a larger $N$ implies a longer search time, allowing perturbations to grow larger, which in turn necessitates a higher $\lambda$ to constrain their size. Finally, forecast lead times were chosen to strike a balance: sufficiently close to the event to ensure forecastability, yet distant enough to allow the introduced perturbations adequate time to evolve.

\begin{center}
\begin{table}[ht]
\makebox[\textwidth]{%
\resizebox{1.0\textwidth}{!}{%
\begin{tabular}{@{}lccccccccccccc@{}}
\hline
Experiment name & $\alpha$ & $\beta$ & $\lambda_{T}$  & $\lambda_{SP}$ & $\lambda_{\delta}$ & $\lambda_{\zeta}$ & $\lambda_{SH}$ & $\lambda_{SCIWC}$ & $\lambda_{SCLWC}$  & Initial Date & $\tau$ & Total integration time & steps\\
\hline
EXP50 & $10^{-9}$ & 20 & 200 & 20 & 2000 & 2000 & 200 & 20 & 20 &  June 21st 2021 & 5 days & 11 days & 50\\
EXP75 & $10^{-9}$ & 10 & 100 & 10 & 1000 & 1000 & 100 & 10 & 10 &  June 21st 2021 & 5 days & 11 days & 75\\
\hline
\end{tabular}
}}
\caption{Parameters used during the optimization process. Each row corresponds to one experiment. The coefficients $\lambda_{T}$, $\lambda_{SP}$, $\lambda_{\delta}$, $\lambda_{\zeta}$, $\lambda_{SH}$, $\lambda_{SCIWC}$, and $\lambda_{SCLWC}$ control the relative weight of the temperature term, the surface pressure term, the divergence term, the vorticity term, the specific humidity term, and the ice and liquid cloud water terms in the loss function. The parameter $\beta$ sets the strength of the temperature objective term. The number of iteration steps differs between the two experiments in order to explore the effect of longer and shorter optimization procedures while all other settings are kept fixed. The quantity $\tau$ denotes the forecast lead time used when computing the loss.}
\label{tab:params}
\end{table}
\end{center}

The optimized simulations are compared to an ensemble run of the event using the stochastic version of NeuralGCM. This ensemble consists of 75 members. Unlike our approach, which perturbs the initial conditions (inputs to the model), the stochastic model introduces perturbations within the learned physics module. As a result, the perturbations are effectively introduced one time step apart. Additionally, our method perturbs surface pressure, which is not perturbed in the stochastic model. More details about the stochastic model can be found in \cite{Kochkov2024}.

\subsection{Initial Condition Perturbations}

Table~\ref{tab:perturb} presents the maximum perturbations applied to the initial conditions, alongside the range of values sampled in the ensemble simulation. The range is computed by finding the max perturbation of all the ensemble members with respect to the mean. Overall, the applied perturbations during optimization remain within or below the variability represented in the stochastic ensemble. Visualizing the perturbations directly is challenging due to their high dimensionality, but their spatial spectrum provides some insight. We present the perturbation spectrum in App.~\ref{APP_spect}.

\begin{center}
\begin{table}[ht]
\makebox[\textwidth]{%
\resizebox{1.0\textwidth}{!}{%
\begin{tabular}{@{}lccc@{}}
\hline
\textbf{Quantity / Experiment} & EXP50 & EXP75  & 75-member ensemble\\
\hline
$\#$ of steps  & 50 & 75 & -\\
Surface Pressure (Pa)  & 0.69 & 0.47 & 0.0\\
Specific Humidity   & $2.34 \times 10^{-3}$  & $1.11 \times 10^{-3}$ & $3.20\times 10^{-3}$\\
Specific Cloud Ice Water Content (kg kg$^{-1}$) & $7.61 \times 10^{-6}$ & $5.19 \times 10^{-6}$ & $4.35\times 10^{-5}$\\
Specific Cloud Liquid Water Content (kg kg$^{-1}$) & $1.50 \times 10^{-5}$ & $2.42 \times 10^{-5}$ & $7.61 \times 10^{-5}$\\
Temperature (K) & 4.83 & 4.99 & 7.60\\
U-component of windspeed (m s$^{-1}$)   & 8.37 & 5.59  & 12.70\\
V-component of windspeed  (m s$^{-1}$)  & 4.77 & 4.12 & 7.94\\
\hline
\end{tabular}
}}
\caption{Maximum perturbations over the full 3D fields for different run sizes compared to the range of perturbations applied on the ensemble run by the stochastic model.}
\label{tab:perturb}
\end{table}
\end{center}

\section{Results}
\label{Sec:Results}

\subsection{NeuralGCM temperature evaluation}

\begin{figure}
    \centering
    \includegraphics[width=\linewidth]{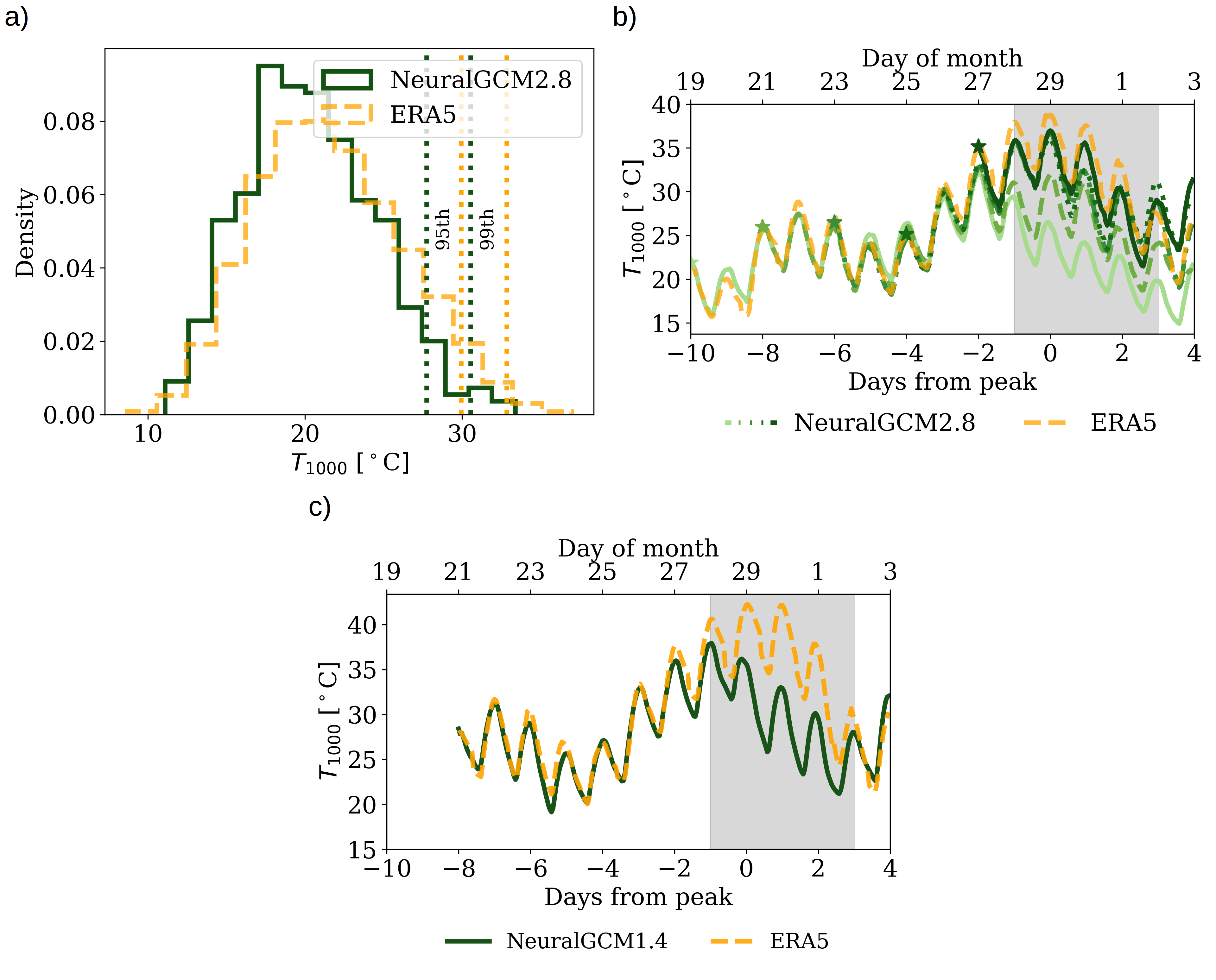}
    \caption{(a) Histograms of 6-hourly temperature values from a 40-year NeuralGCM simulation (green) and from ERA5 (orange) reanalysis over the study domain outlined in Figure~\ref{fig:timeseries_main_var}. Dashed line indicates the 95th and 99th percentiles. (b) Time series of temperature forecasts at 1000-hPa from NeuralGCM with 10-, 6-, 4-, and 2-day lead times (green colored lines) compared with ERA5 reanalysis data (orange line) for the PN2021 heatwave. Grey area highlights the targeted time range for the optimization process.}
    \label{fig:ERA_EVAL_28}
\end{figure}

We first evaluate the ability of the NeuralGCM2.8 model to simulate a temperature summer climate (June, July, August) compared to ERA5. Figure~\ref{fig:ERA_EVAL_28}a) presents the 6-hourly temperature distribution for a NeuralGCM2.8 simulation and the ERA5 data over the 1981-2020 40-year period, averaged within the domain of interest (highlighted in the blue box in Fig.~\ref{fig:timeseries_main_var}). The NeuralGCM2.8 simulation closely approximates the ERA5 distribution, demonstrating its ability to reproduce key statistical characteristics of the temperature distribution in this region. We highlight the 95th and 99th percentile values for both the model and ERA5. We note that the model has slightly colder hot extremes than ERA5.

Next the ability of the NeuralGCM2.8 model to forecast the PN2021 heatwave against the ERA5 reanalysis data is evaluated. Figure~\ref{fig:ERA_EVAL_28}b) shows NeuralGCM2.8 forecasted surface temperatures during the PN2021 heatwave for five lead times ranging from 10 to 2 days.
At a 10-day lead time, the NeuralGCM2.8 predictions follow closely the ERA5 data until about day 8, where they deviate leading to a lack of heatwave and extreme temperatures. At an 8-day lead time, the simulation substantially enhances temperature during the heatwave period but still shows large underestimations of peaks intensities, with differences of about $6^\circ$C. At 2-, 4-, and 6-day lead times, the NeuralGCM2.8 model captures well the general pattern of temperature variations shown by ERA5, including the occurrence of very high temperatures during the heatwave event. However, most forecasts underestimate the peak magnitude compared to ERA5 by a few degrees Celsius, particularly during the days after the peak of the event. 

This underestimation of the extreme heat, to our knowledge, is due to two factors: 1) there seems to be a dependence on capturing the extreme with the coarseness of the model, when we increase the resolution to the $1.4^\circ$ model, the prediction quality improves (see Fig.~\ref{fig:ERA_EVAL_28}C)). 2) other studies have evaluated the ability of simulating extreme heatwave storylines and found that the model lacking processes, such land-surface feedbacks led to under representation of extreme \citep{duan2024aibasedclimatemodelevaluation}.

\subsection{Optimizing extreme temperatures}
 
\begin{figure}
    \centering
    \includegraphics[width=1.0\linewidth]{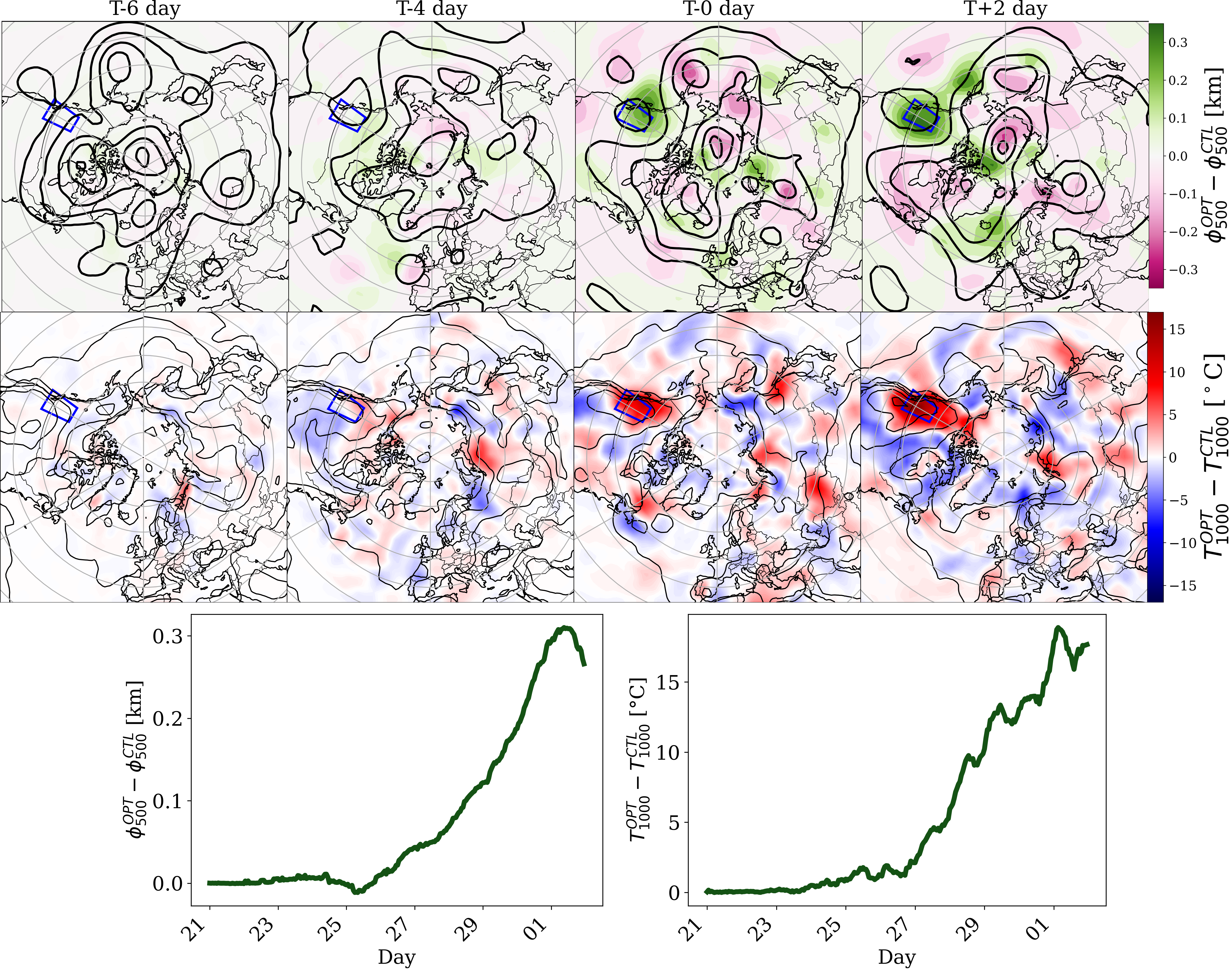}
    \caption{Top row: Evolution of the difference in 500-hPa geopotential height ($\Delta Z$, in km) between the optimized simulation and the control run for EXP75. Black contours (optimized run) outline the amplified Rossby wave pattern, with deeper troughs and higher ridges compared to the control. Middle row: The difference in 1000-hPa temperature (${}^\circ$C) between the optimized and control simulations. Bottom row: Th difference between the 500-hPa geopotential height and 1000-hPa temperature averaged over the target domain.}
    \label{fig:hemisphere_diff}
\end{figure}

We optimize the initial conditions of the NeuralGCM model starting from June 21st, 2021 (corresponding to a lead time of 8 days to the peak of PN2021; see Sect.~\ref{Sec:Methods} for details) and run the simulation forward for 11 days. An 8-day lead time strikes a balance between two requirements: keeping the event within the model’s predictable window and allowing enough time for small perturbations to develop. The optimization is performed using gradient descent over 75 steps to solve eq.~\ref{eq:opt_problem}, targeting the last 5-days of the event (see grey shaded area in Fig.~\ref{fig:ERA_EVAL_28}). The full set of parameters used in the optimization is provided in Tab.~\ref{tab:params}. Figure~\ref{fig:hemisphere_diff} shows the differences in 500-hPa geopotential height ($\phi_{500}$; top row) and the 1000-hPa temperature ($T_{1000}$; bottom row) between the optimized (OPT) and control (CTL) trajectories. In addition, the $\phi_{500}$ and $T_{1000}$ of the optimized simulation are shown in dark contour lines. The early day conditions (T-6 day) show minimal differences, with anomalies amplifying progressively as we get closer to the peak day. Positive and negative differences in $\phi_{500}$ are generally observed in association with ridges and troughs, respectively, indicating that the optimized simulation amplifies the hemispheric wave amplitudes. Specifically, over the targeted region, there is a clear increase in $\phi_{500}$ and $T_{1000}$, with the largest increases centered over the targeted region.

\begin{figure}
    \centering
    \includegraphics[width=1.0\linewidth]{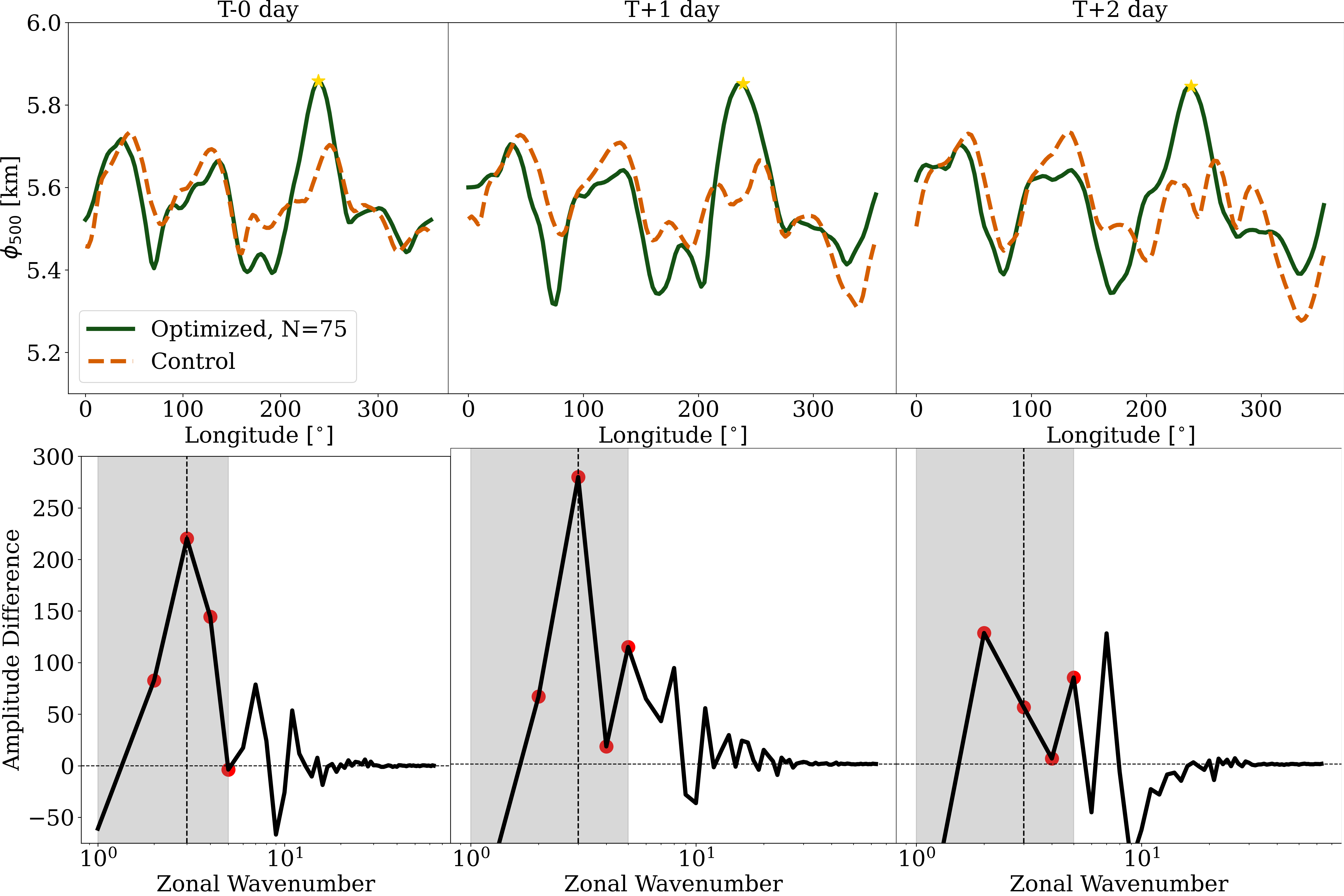}
    \caption{Top row: cross section of 500-hPa geopotential along a fixed latitude for EXP75. Bottom row: the amplitude difference of the Fourier spectrum, including wavenumbers 1–5 highlighted in grey. Red dots highlight wavenumbers 2-5.}
    \label{fig:ross_wave}
\end{figure}

We examine the 500-hPa geopotential height along a fixed latitude (latitude = $57.2^{\circ}$) in Fig.~\ref{fig:ross_wave}. The wave patterns produced by the optimized simulation are compared to those from the control simulation, along with their respective spectral characteristics during the last three days of the event. Notably, the geopotential height near the heatwave region is significantly higher in the optimized simulation compared to the control one. Both the control and optimized simulations show signs of a persistent wavenumber three wave. In the spectral amplitude, the largest differences in spectral amplitude occur for wavenumbers 2–5, which are typically associated with heatwave events. Specifically, the largest differences are observed at wavenumber 3, followed by 2 and 4, where the optimized simulation exhibits greater power than the control simulation. While some differences are also present at higher wavenumbers, their magnitude is substantially smaller.

\begin{figure}
    \centering
    \includegraphics[width=0.75\linewidth]{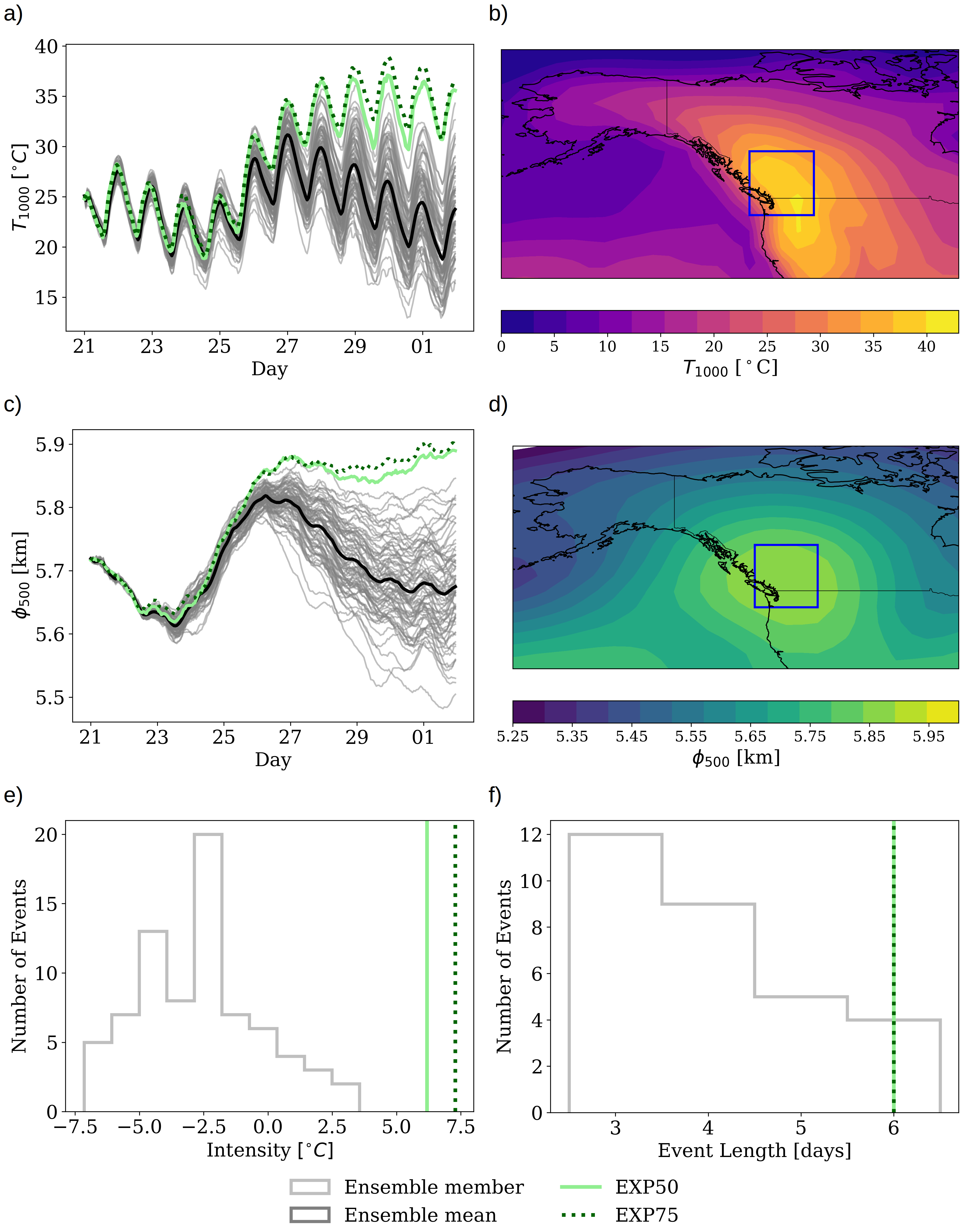} 
    \caption{(a) Time series of 1000-hPa temperature for two optimized trajectories (50 and 75 steps) and a 75-member ensemble with its mean. (b) Spatial map of average temperature anomalies from the 75-step run. (c) Time series of 500-hPa geopotential height for the same set of simulations. (d) Spatial map of 500-hPa geopotential height anomalies during the event period. (e, f) Time series of heatwave intensity (defined by Eq.~\ref{eq:hw_def}) and duration for the ensemble and optimized cases.}
    \label{fig:timeseries_main_var}
\end{figure}

The optimization process relies on gradient descent (see Sect.~\ref{Sec:Methods} for details), which requires choosing the number of gradient descent steps. Figure~\ref{fig:timeseries_main_var} shows hourly time series of $T_{1000}$ (Figure~\ref{fig:timeseries_main_var}a) and $\phi_{500}$ (Figure~\ref{fig:timeseries_main_var}c) for two optimized trajectories with $N=50$ (EXP50) and $N=75$ (EXP75) steps, alongside a 75-member ensemble and its mean, all initialized 8 days before the peak of the event. Notably, the $T_{1000}$ time series (Fig.~\ref{fig:timeseries_main_var} a) reveals that both optimized trajectories attain values beyond the range exhibited by any individual ensemble member. In other words, the proposed method allows us to find extreme temperature values that are more extreme than those found using a 75-member ensemble using only 50 iterations (a 33\% reduction in computational cost relative to generating the 75-member ensemble, calculated as $(75-50)/75$). Notably, this more efficient 50-step optimization run produces a trajectory more extreme than any member of the 75-member ensemble. Specifically, the trajectory from 50 steps reaches a peak temperature of $37.0^{\circ}$C, while the trajectory after 75 steps attains $38.9^{\circ}$C. Compared to the mean of the ensemble, we reach anomalies of $14.0^{\circ}$C. For the 500-hPa geopotential height (Fig.~\ref{fig:timeseries_main_var}c), both optimized trajectories show similarly elevated values, once again exceeding the range spanned by the 75-member ensemble. Importantly, the trajectory from the 75-step optimized run maintains a more sustained increase of the 500-hPa geopotential height compared to that from the 50-step run. The measured intensity (Fig.~\ref{fig:timeseries_main_var}e) and length of the event (Fig.~\ref{fig:timeseries_main_var}f) are increased in both optimized runs compared to the ensemble mean. Both optimized runs produce a 6-day-long heatwave event, differing only in intensity, with the 75-step run having a $1.0^\circ$C higher intensity than the 50-step run. Notably, both optimized solutions exceed the intensities spanned within the 75-members ensemble. Figure \ref{fig:timeseries_main_var}b,d present the temperature and geopotential fields from the 75-iteration run. The temperature pattern features a maximum over the targeted region, albeit slightly to the south, while the 500-hPa geopotential height field exhibits an anticyclone directly overhead. The resulting fields from the optimized solution are consistent with what is seen in ERA5 data for the PN2021 event as can be seen in App.~\ref{APP:ERA5}.

\subsection{Sensitivity of other variables}
\begin{figure}
    \centering
    \includegraphics[width=0.75\linewidth]{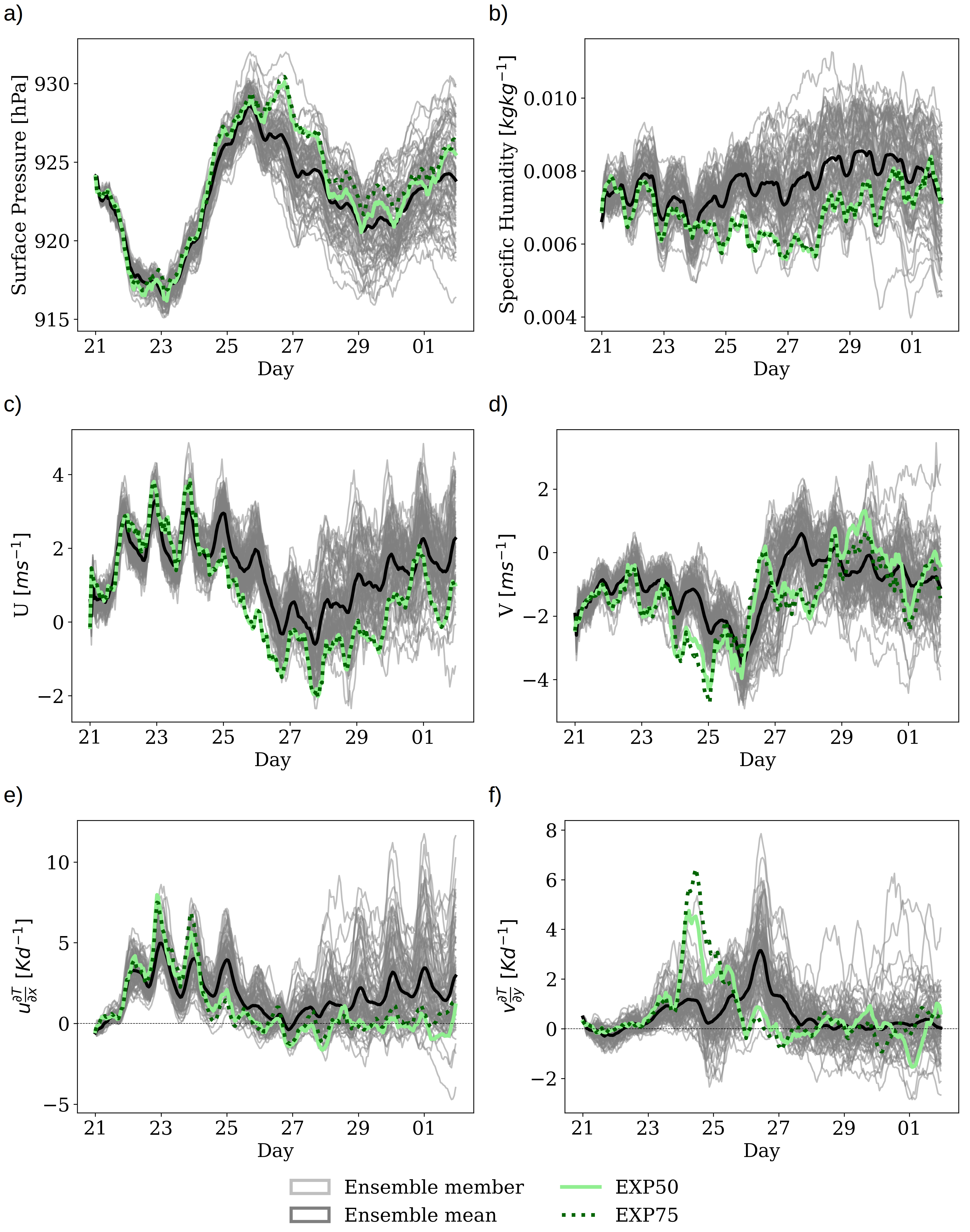}
    \caption{Time series of (a) surface pressure, (b) near-surface specific humidity, (c) U-component of wind, (d) V-component of wind, (e) and (f) temperature advection for each component at 1000-hPa . Data from the optimized trajectory are shown alongside the individual ensemble members (in grey) and the ensemble mean (thick black line).}
    \label{fig:timeseries_other_var}
\end{figure}

Figure \ref{fig:timeseries_other_var} shows optimized trajectories for near-surface wind (zonal U and meridional V components), specific humidity, surface pressure, and temperature advection. In the optimized run, the U wind lies consistently at the lower end of the ensemble spread during the event, and specific humidity likewise tracks near the lower end. Such concurrent reductions in near-surface wind speeds and humidity are consistent with the physical mechanisms that underlie heatwave intensification. In contrast, the V wind and sea-level pressure exhibit only slight positive anomaly above the ensemble mean. All optimized trajectories remain entirely within the bounds defined by the non-optimized ensemble members. While the variables are within range of the ensemble envelope (i.e., not extreme), there might be a confluence of factors that leads to the extreme.

\subsection{Sensitivity to NeuralGCM resolution}
\label{SEC:Sensitivity}
\begin{figure}
    \centering
    \includegraphics[width=0.75\linewidth]{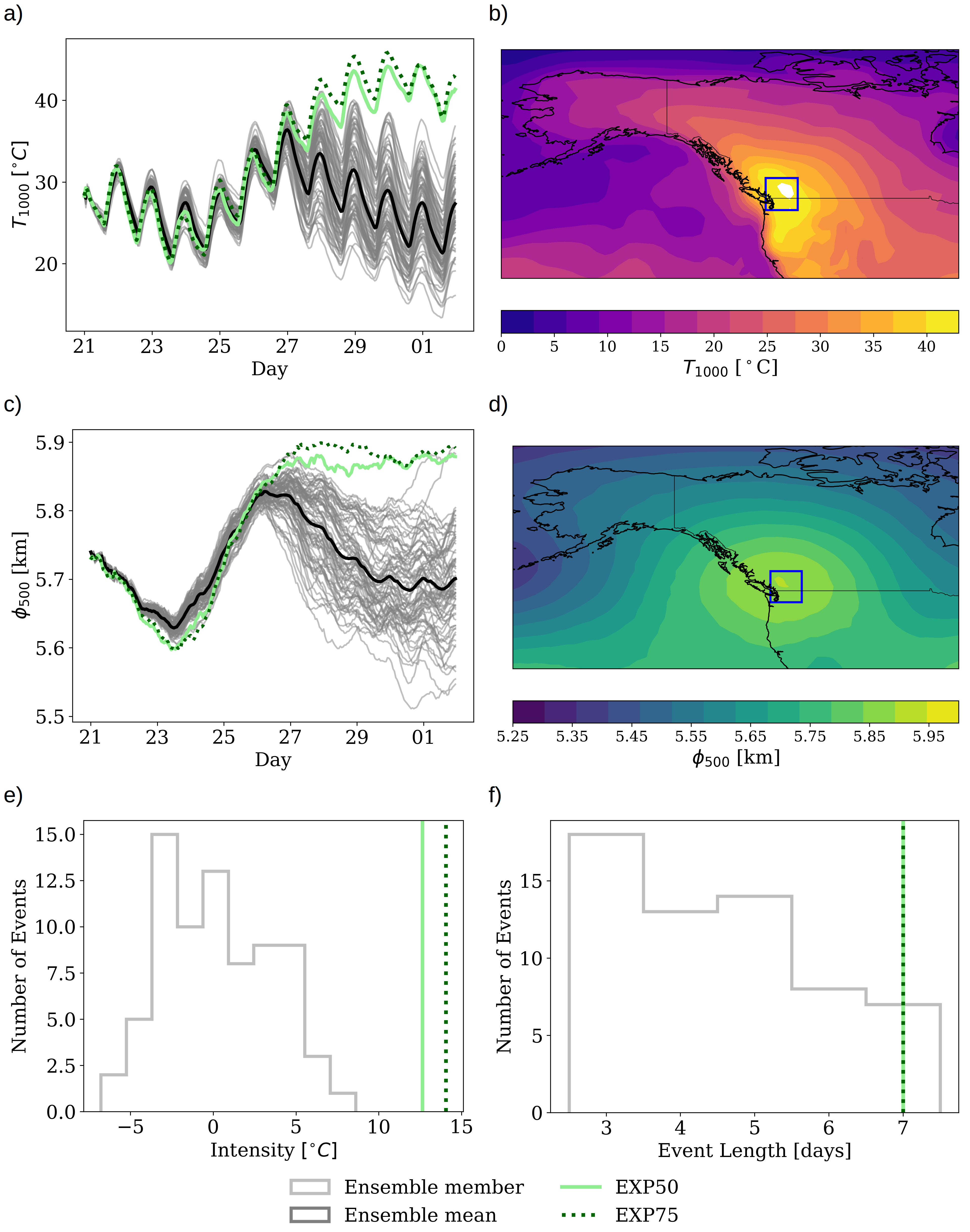} 
    \caption{Same as Fig.~\ref{fig:timeseries_main_var} except with $1.4^{\circ}$ resolution model. (a) Time series of 1000-hPa temperature for two optimized trajectories (50 and 75 steps) and a 75-member ensemble with its mean. (b) Spatial map of average temperature anomalies from the 75-step run. (c) Time series of 500-hPa geopotential height for the same set of simulations. (d) Spatial map of 500-hPa geopotential height anomalies during the event period. (e, f) Time series of heatwave intensity (defined by Eq.~\ref{eq:hw_def}) and duration for the ensemble and optimized cases.}
    \label{fig:timeseries_main_var_hr}
\end{figure}

To test how resolution affects our optimization, we reran the optimization problem on NeuralGCM at a finer $1.4^\circ$ resolution (Figure \ref{fig:timeseries_main_var_hr}). The same parameters as in Table.~\ref{tab:params} is used for this set of experiments. In the unperturbed control, the high-resolution ensemble mean reduces the warm bias against ERA5 and more accurately captures the peak and decay of the PN2021 heatwave. Once optimized, the $1.4^\circ$ run again produces peak surface‐temperature anomalies that exceed the 75‐member ensemble maximum, and 500-hPa geopotential‐height anomalies that surpass the control even more than the $2.8^\circ$ case. These enhanced temperature and geopotential height anomalies persist through the extended target period, demonstrating that the optimization delivers sustained extremes even at higher resolution.

\section{Discussion}
\label{Sec:Discussion}

Our findings demonstrate that differentiable climate models, exemplified by NeuralGCM, offer a powerful tool for constructing extreme heatwave storylines through gradient-based optimization of initial conditions. By perturbing initial conditions, we identified alternative trajectories with slightly different synoptic-scale conditions that amplify the PN2021 heatwave intensity by $3.7^\circ$C according to NeuralGCM. These results align with prior studies linking extreme heat to persistent blocking patterns \citep{Screen2014,ThePacificNorthwestHeatWave}. Specifically, the optimized geopotential height anomalies and spectrum reflect enhanced blocking dynamics with an amplification of wavenumbers 1-5. The resulting temperature and geopoentatial increase has realistic features when comparing the to ERA5 data as can be seen in App.~\ref{APP:ERA5}.

While NeuralGCM resolves large-scale dynamics, its omission of land-atmosphere feedbacks (e.g., soil moisture \citep{duan2024aibasedclimatemodelevaluation}) likely results in a conservative estimate of heatwave amplification. For instance, soil moisture–temperature coupling is known to cause stronger heatwave persistence (e.g., \citep{Suarez-Gutierrez2020, White2023}), implying that the model might underestimate extremes when such feedbacks are neglected. Additionally, the model's coarse horizontal resolution ($2.8^\circ$) introduces biases in capturing localized extreme conditions associated with the PN2021 event. As show with the higher resolution simulation, the use of finer grids ($1.4^\circ$) allows for more accurate estimates of the extreme temperatures of the event although the main dynamical changes remain similar to the coarse resolution. In addition, while we have chosen to use NeuralGCM for this study, the method could be applied to any model which had automatic differentiation implemented. This includes all existing purely data-driven models, including GraphCast, Pangu-weather, FourCastNet, and FuXi. While data-driven models could provide faster predictions at higher resolution, their dual-initial-condition requirement introduces ambiguity about finding optimal initial conditions. For instance, \cite{vonich2024predictabilitylimit2021pacific} optimized both inputs for GraphCast to reconstruct the 2021 heatwave, but this approach demands simultaneous perturbation of two distinct states. Validation against models like Pangu demonstrated that results remained consistent despite this added complexity, suggesting robustness in the dual-input framework. However, NeuralGCM’s hybrid design simplifies the workflow by requiring only a single initial condition. In addition, \cite{https://doi.org/10.1029/2023GL105747} show that data-driven forecast models tend to poorly represent small perturbations, often filtering them out, which could impact the method. The extent to which this affects hybrid models, which has a dynamical core, such as NeuralGCM remains unclear.

The optimization process involves making several decisions and setting specific parameters. While we do not present the results here, we have explored a limited subset of the broader hyperparameter space—specifically, the learning rate ($\alpha$) and the loss function parameters ($\beta$,$\vec{\lambda}$). We found that, for the loss function parameters defined in Sect.~\ref{Sec:Methods}, a large learning rate induces instability, causing substantial perturbations without a corresponding increase in temperature and in some cases, triggering numerical instabilities that caused the simulation to fail. To ensure stability, $\alpha$ must be on the order of $10^{-9}$ or smaller. This stability condition varies in a nonlinear and nontrivial manner with changes in $\vec{\lambda}$. Furthermore, the consistency of the results across the EXP50 and EXP75 experiments and the simulations at two different resolutions—all of which yield trajectories more extreme than the 75-member stochastic ensemble—suggests that the optimized perturbations are not simply initialization artifacts. However, a systematic quantification of the sensitivity to the initial state and a thorough exploration of the hyperparameter space lie beyond the scope of this study. Fine-tuning the parameters might allow for improved efficiency in computational cost, where a more extreme solution is found with a reduced number of steps. In addition we note that we have chosen to optimize the 1000-hPa level but, NeuralGCM utilizes $\sigma$-coordinates. Over regions with significant elevation, such as the Canadian Rockies, the 1000-hPa geopotential surface is often below ground level. Using the 1000-hPa temperature ($T_{1000}$) can therefore yield physically inconsistent values when optimizing for near-surface extreme events. To ensure the optimized initial conditions lead to physically meaningful and surface-relevant extreme temperatures across the entire domain, we analyze in App.~\ref{APP:850} the 850-hPa temperature ($T_{850}$).

To evaluate the robustness and physical realism of the NeuralGCM-optimized initial conditions, these perturbations should be tested in a conventional numerical weather prediction model (e.g., Environment and Climate Change Canada's model \cite{buehner_implementation_2015}). Such cross-model validation would reveal the universality of the results and help isolate NeuralGCM-specific biases. Additionally, running these scenarios in a fully physical model would explicitly account for land–atmosphere interactions and feedbacks, and assess whether the extreme trajectories persist under more detailed dynamics and physics.

Our method focuses on optimizing initial conditions, assuming the underlying model physics (whether learned or explicit) are fixed and skillful. An alternative approach could involve optimizing model parameters themselves  (as done for example by \cite{alet2025skillfuljointprobabilisticweather} to generate ensembles), though this would require careful regularization to ensure the resulting model remains physically plausible.

The computational efficiency of the ML and hybrid models coupled with their differentiable properties, opens avenues for exploring extreme events—from heatwaves to precipitation extremes and compound disasters. For example, a similar optimization problem could be formulated for StormCast \citep{pathak2024kilometerscaleconvectionallowingmodel} to allow us to search for extreme precipitation events in an emulated regional climate model. One could also formulate the loss function such that the large deviation theory rate function is minimized, leading to ``typical'' trajectories of extremes \citep{10.1063/1.5084025, PhysRevX.13.041044}. We could also envision loss functions with hard constraints on the perturbation which impose conservation laws as opposed to simply imposing small perturbations.

\conclusions  
\label{Sec:Conclusions}
We introduce a differentiable-storyline framework that leverages automatic differentiation in hybrid climate models to directly optimize initial conditions and generate physically coherent extreme-heatwave trajectories at a fraction of the computational cost of traditional ensemble methods. For example, our 50-step optimization run produced a more extreme event than any member of a 75-member ensemble, while using 33\% less computational resources than it took to generate that ensemble. When applied to the PN2021 heatwave, our approach produces intensifications of nearly $3.7^\circ$C by isolating high-impact circulation patterns, specifically enhanced blocking and Rossby-wave, demonstrating both its dynamical fidelity and efficiency. While this proof-of-concept focuses on NeuralGCM and a single case study, the optimization paradigm is agnostic to model architecture and event type, offering a transformative tool for rapid, process-based risk assessment of diverse climate extremes in a warming world.

\codeavailability{\href{https://github.com/timwhittaker/ExtremeStorylines}{https://github.com/timwhittaker/ExtremeStorylines} (\href{https://doi.org/10.5281/zenodo.15649393}{https://doi.org/10.5281/zenodo.15649393})} 

\dataavailability{Data availability: All data generated using the code and ERA5 data used in this study is openly available from the Copernicus Data Store (\href{https://doi.org/10.24381/cds.143582cf}{https://doi.org/10.24381/cds.143582cf}) \citep{ERA5}.} 

\appendix

\section{PN2021 ERA5}
\label{APP:ERA5}
\begin{figure}[ht]
    \centering
    \includegraphics[width=\linewidth]{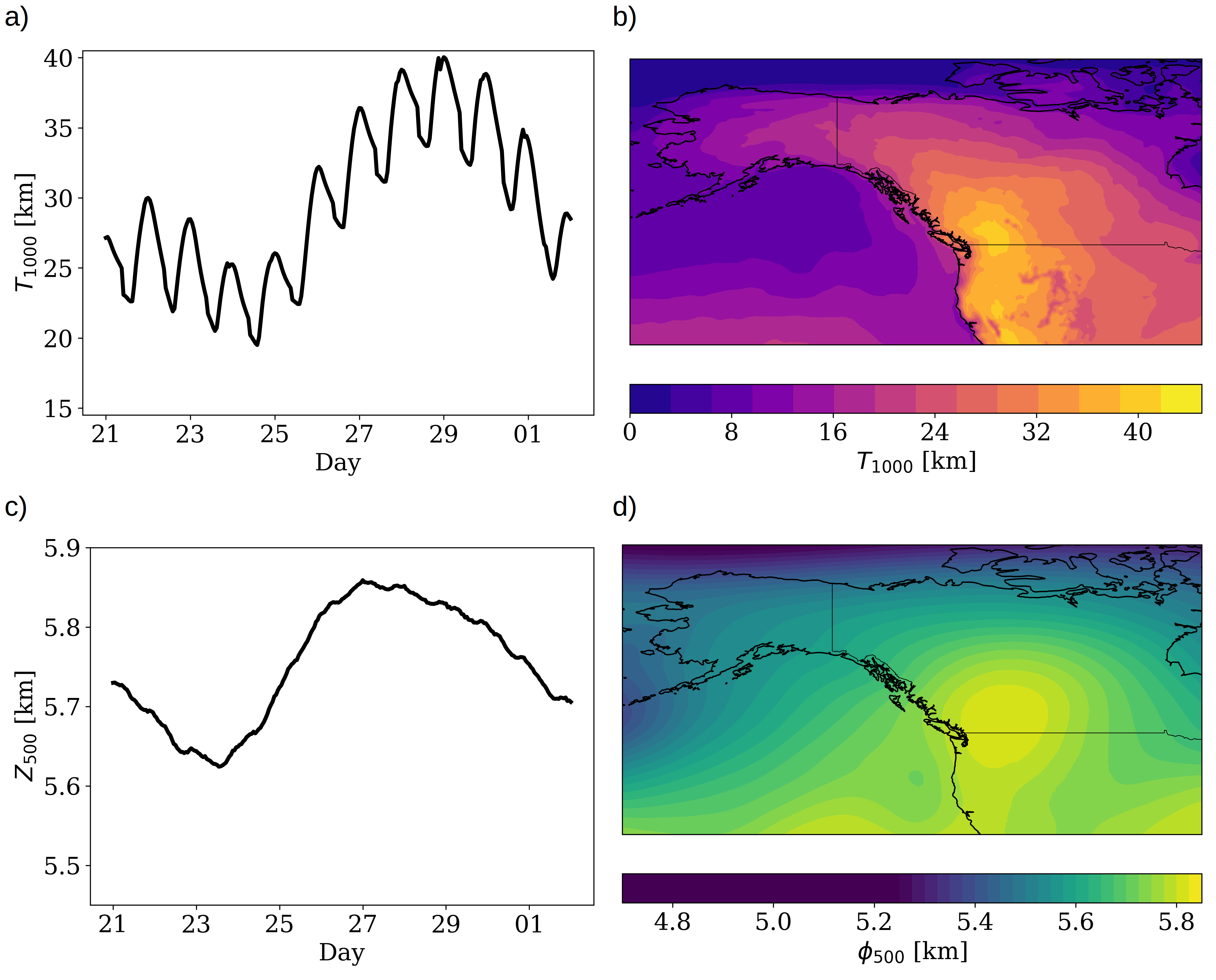}
    \caption{(a) Time series of 1000-hPa temperature for ERA5 for the PN2021. (b) Spatial map of average temperature anomalies from the ERA5 data. (c) Time series of 500-hPa geopotential height for ERA5. (d) Spatial map of 500-hPa geopotential height anomalies during the event period.}
    \label{fig:pn2021_era5}
\end{figure}

\begin{figure}[ht]
    \centering
    \includegraphics[width=\linewidth]{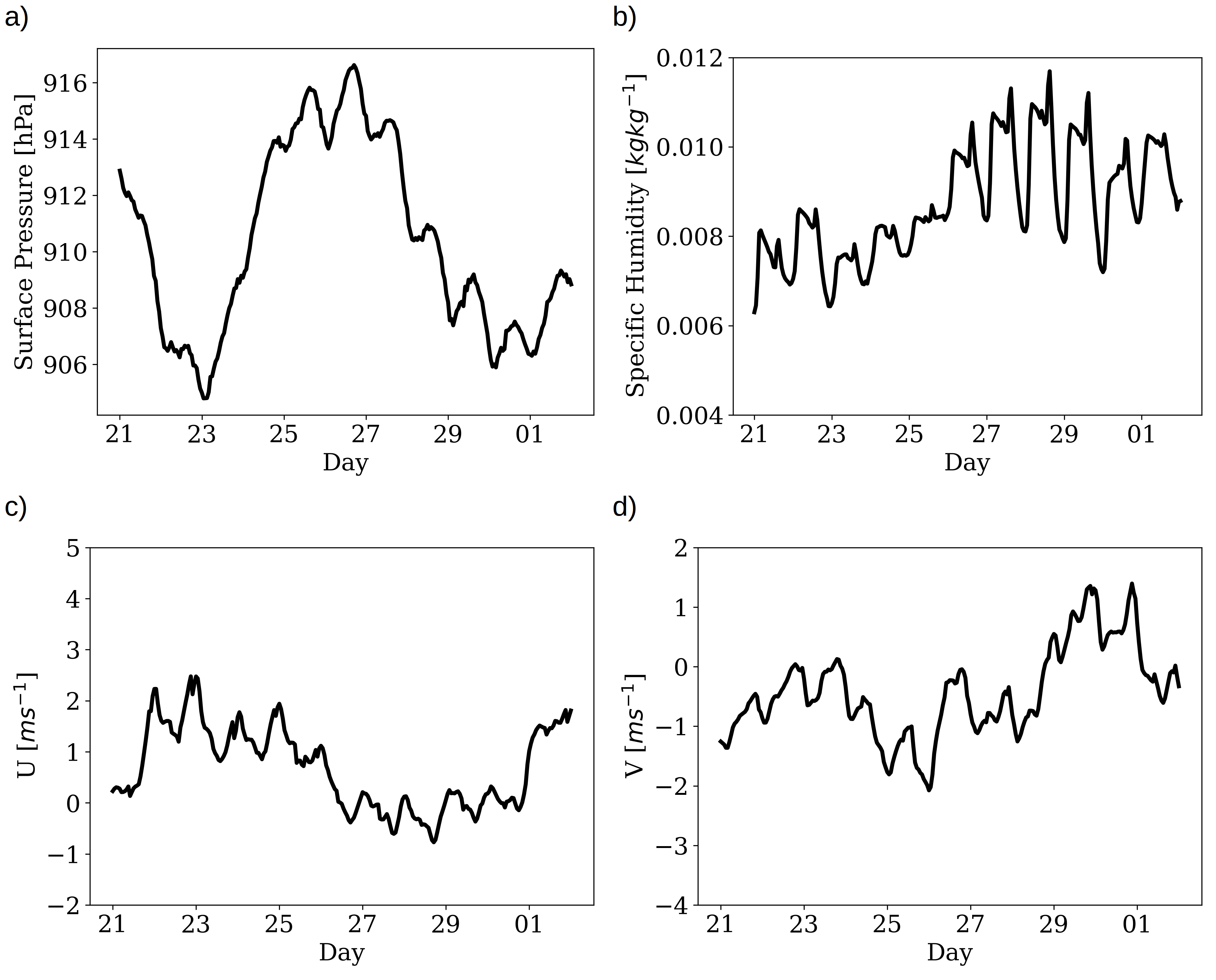}
    \caption{Equivalent of figure 5 using ERA5 data. Time series of (a) surface pressure (b) near-surface specific humidity (c) U-component of wind and (d) V-component of wind at 1000-hPa.}
    \label{fig:fig5_era}
\end{figure}

For reference, Figure \ref{fig:pn2021_era5} presents the atmospheric fields observed during the 2021 heatwave from the ERA5 reanalysis dataset. The figure displays the 1000-hPa temperature ($T_{1000}$) and the 500-hPa geopotential height ($Z_{500}$), characterizing the event. This illustrates the large-scale structure, such as the high-amplitude ridge in $Z_{500}$, associated with the event's formation. Fig.~\ref{fig:fig5_era} shows the other variables as in Fig.~\ref{fig:timeseries_other_var}. The specific humidity, and winds are within the envelop of the NeuralGCM ensemble. The surface pressure on the other hand has a positive bias in NeuralGCM likely due to the representation of the surface in the coarse NeuralGCM model.

\section{Perturbation Spectrum}
\label{APP_spect}
\begin{figure}[ht]
    \centering
    \includegraphics[width=\linewidth]{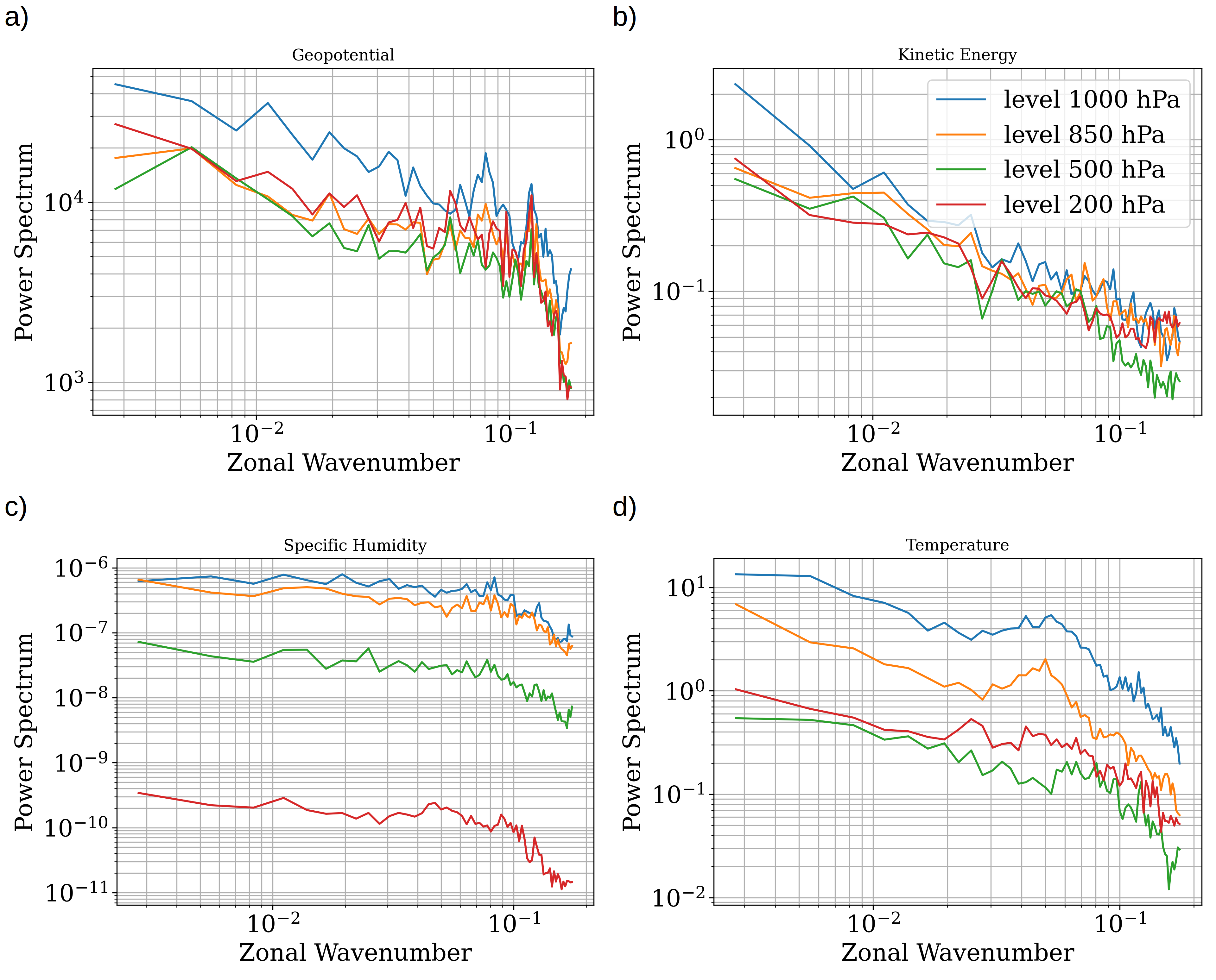}
    \caption{Spectrum of the three-dimensional perturbations at selected vertical levels for the geopotential, temperature, kinetic energy and specific humidity.}
    \label{fig:perturbation_spectrum}
\end{figure}

The perturbations introduced in the optimized runs are fully three-dimensional and span all horizontal and vertical levels. Due to this high dimensionality, it is challenging to visualize their full structure directly. To provide some insight into their characteristics, we show the spatial spectrum of the perturbations at selected vertical levels in Figure~\ref{fig:perturbation_spectrum}. This representation highlights the dominant spatial scales of the perturbations across the domain. A more detailed analysis of their full spatial structure could be informative, but is beyond the scope of this work.

\section{Temperature at 850-hPa}
\label{APP:850}
NeuralGCM utilizes $\sigma$-coordinates. Over regions with significant elevation, such as the Canadian Rockies, the 1000-hPa geopotential surface is often below ground level. Using the 1000-hPa temperature ($T_{1000}$) can therefore yield physically inconsistent values when optimizing for near-surface extreme events. To ensure the optimized initial conditions lead to physically meaningful and surface-relevant extreme temperatures across the entire domain, we analyze the 850-hPa temperature ($T_{850}$). 

\begin{figure}[ht]
    \centering
    \includegraphics[width=\linewidth]{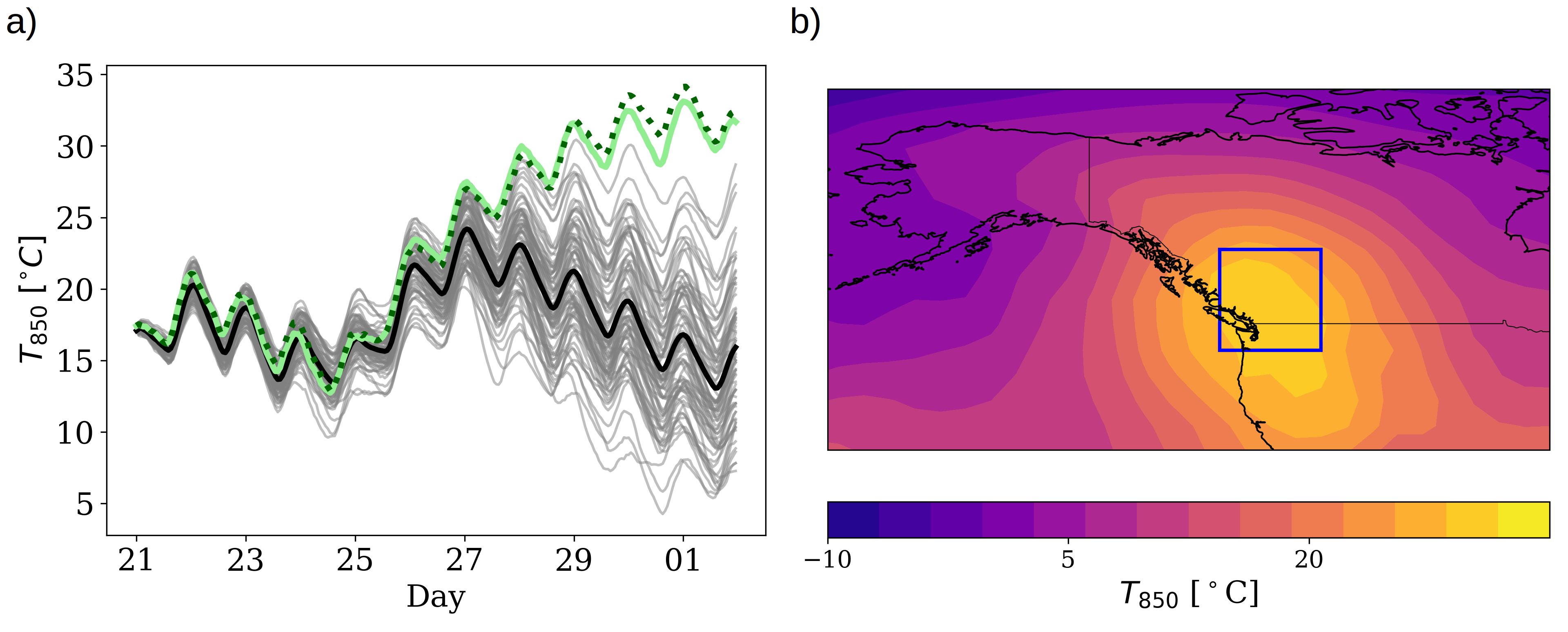}
    \caption{(a) Time series of 850-hPa temperature for EXP50, EXP75 and the ensemble. (b) Spatial map of average temperature anomalies from the EXP75 data.}
    \label{fig:T850}
\end{figure}

Fig.~\ref{fig:T850} presents the $T_{850}$ fields for the optimized extreme events. Despite the optimization targeting $T_{1000}$ the temperature at the 850-hPa level still exhibits a clear increase, exceeding the values observed in the ensemble simulations. We note that the magnitude of the anomaly found at 850-hPa is smaller than the maximum value achieved at the 1000-hPa level.

\authorcontribution{T.W. and A.D.L conceptualized the method. T.W. designed and performed the numerical experiments. T.W. and A.D.L prepared the manuscript with equal contribution.} 

\competinginterests{No competing interests} 

\begin{acknowledgements}
T.W. would like to thank Robin Noyelle, and Eric Vanden-Eijnden for discussions which lead to this project. T.W. would like to thank Seth Taylor and Tangui Picart for discussions and comments on the draft. The authors would like to thank Frédérik Toupin, Katja Winger, and Francois Roberge for maintaining a user-friendly local computing facility. This research was enabled in part by support provided by Calcul Québec (calculquebec.ca) and the Digital Research Alliance of Canada (alliancecan.ca). 
\end{acknowledgements}

\newpage
\bibliographystyle{copernicus}
\bibliography{sn-bibliography}

\end{document}